\documentclass[11pt,titlepage]{article}

\usepackage{setspace}
\usepackage{amsmath}

\usepackage{graphicx}

\usepackage[round,numbers,sort&compress]{natbib}
\newcommand{\msd}{\ensuremath{\langle r^{2} \rangle}}
\newcommand{\msdt}{\ensuremath{\langle r^{2}(t) \rangle}}
\newcommand{\sitj}{\ensuremath{\hat{s}_j(t)}}
\newcommand{\siz}{\ensuremath{\hat{s}(0)}}
\newcommand{\sit}{\ensuremath{\hat{s}(t)}}

\title{Fluorescence Correlation Spectroscopy analysis of segmental dynamics in Actin filaments}

\author{Anne~Bernheim-Groswasser \\
    Chemical Engineering Department, Reimund Stadler Minerva Center\\
     and Ilse Kats Center for Nanoscience, \\
    Ben-Gurion University, Beer-Sheva, Israel
    \and Roman Shusterman\\
    Physics Department \\
    Ben-Gurion University, Beer-Sheva, Israel
    \and Oleg Krichevsky \thanks{
                    Corresponding author. Address:
                    Physics Department,
                    Ben-Gurion University,
                    Beer-Sheva, 84105 Israel,
                    Tel.:~(972)8647-2123, Fax:~(972)8647-2904, Email:~okrichev@bgu.ac.il} \\
    Physics Department and Ilse Kats Center for Nanoscience,\\
    Ben-Gurion University, Beer-Sheva, Israel}

\date{}

\begin{document}

\maketitle

\abstract{We adapt Fluorescence Correlation spectroscopy (FCS) formalism to the
studies of the dynamics of semi-flexible polymers and derive expressions relating
FCS correlation function to the longitudinal and transverse mean square
displacements of polymer segments. We use the derived expressions to measure the
dynamics of actin filaments in two experimental situations: filaments labeled at
distinct positions and homogeneously labeled filaments. Both approaches give
consistent results and allow to measure the temporal dependence of the segmental
mean-square displacement (MSD) over almost five decades in time, from $\sim
40\mu$s to $\sim 2$s. These noninvasive measurements allow for a detailed
quantitative comparison of the experimental data to the current theories of
semi-flexible polymer dynamics. Good quantitative agreement is found between the
experimental results and theories explicitly accounting for the hydrodynamic
interactions between polymer segments.

\emph{Key words:}polymer dynamics, fluorescence correlation spectroscopy,
F-Actin, semi-flexible polymers }

\clearpage

\section*{Introduction}
Living cells have remarkable mechanical properties which
enable them to move, divide and respond to external stresses. These properties
are mainly attributed to the dynamical characteristics of the cell cytoskeleton,
which is a complex three-dimensional network of protein filaments mostly
comprised of F-actin and microtubules. Both types of filaments are the
polymerized forms of monomeric protein subunits: globular actin (G-actin) and
tubulin, respectively. The cytoskeleton derives its strength from the elastic
properties of these biopolymers, which, unlike synthetic polymers, are
characterized by a high bending rigidity.

The polymer rigidity is described by a persistence length $l_p$ above which
thermal fluctuations can efficiently bend the polymer. Semi-flexible polymers
such as F-actin and microtubules have a long persistence length:
$l_p\sim$17$\mu$m for F-actin \citep{Gittes93} and several millimeters for
microtubules, orders of magnitude larger than those of synthetic polymers. The
large rigidity of these biological polymers enables us to experimentally address
the fundamental questions of polymer dynamics at the sub-persistence length
scales. The dynamics and the mechanical properties of F-actin and microtubule
networks were studied by different optical techniques, such as dynamic light
scattering \citep{Sackmann92}, fluorescence imaging \citep{Kas94, Amblard02},
diffusive wave spectroscopy (DWS) and microrheology \citep{Weitz99, Sackmann00,
Caspi98, Wirtz98, Weitz04}.

One of the most detailed features of polymer dynamics accessible in experiments
is the kinetics of monomer motion measured by the temporal dependence of
monomer's MSD {\msdt}. For length scales below the persistence length, the
monomer displacements are anisotropic with a major contribution coming from the
transverse modes (displacements perpendicular to the polymer contour) and only a
minor contribution from longitudinal modes. Theories \citep{AllegraGanazzoli81,
FargeMaggs93, Harnau96, Harnau97, Granek97, KroyFrey97, Morse98, MacKintosh98}
predict that the kinetics of transverse motion, and thus the overall monomer's
MSD, should follow the power law dependence of $\msd\propto t^{3/4}$.

A feature related to the monomer MSD, the time-dependence of the longitudinal
fluctuations was obtained by measuring the end-to-end distance of individual
actin filaments visualized by fluorescence video-microscopy \citep{Amblard02}.
Although the data are consistent with theoretically predicted dependence, the
temporal resolution of video-microscopy limits the measurements to time scales
larger than 80ms \citep{Amblard02} and thus the overall range of $\msd\propto
t^{3/4}$ dependence spans over one order of magnitude in time only.

A wide range of time scales was assessed using DWS and microrheology of
micron-sized beads inserted into the F-actin mesh \citep{Weitz99, Wirtz98}. The
$\msd\propto t^{3/4}$ scaling was observed from $\sim$10$\mu$s to $\sim$10ms.
However, it is not quite clear how the MSD of a bead is related to the actual
monomers' displacements. While the beads' motion depends on the properties of the
mesh, the beads themselves may affect the dynamics of the filaments, e.g. via
their large friction coefficient \citep{Weitz99}. In particular, the monomer
kinetics measured by fluorescence video-microscopy \citep{Amblard02} is two
orders of magnitude faster than the motion of the beads measured by DWS
\citep{Weitz99}.

Here we present a new non-invasive approach to measure the monomer dynamics in
stiff filaments: the filaments are tagged with fluorescent labels and the
segmental dynamics of the filaments is then followed with Fluorescence
Correlation spectroscopy (FCS) technique. We show that FCS correlation function
is directly related to the temporal dependence of monomers' mean-square
displacement (MSD) $\msdt$. The measurement of the FCS correlation function
allows us to obtain the kinetics of F-actin monomer motion over a wide range of
time scales, from 40$\mu$s to 2s. Previously, the same method was used to study
monomer dynamics in DNA polymers \citep{Shusterman04, Radler03}. However, while
the standard formulas of FCS can be applied to double-stranded DNA, new
expressions relating FCS correlation function to monomer MSD have to be derived
for stiff filaments such as F-actin and used to analyze experimental data. The
main reason for that is the large difference in the persistence lengths of DNA
and F-actin. In general, the dynamics of a semiflexible polymer is anisotropic:
the segmental motions transverse to the filament are larger than longitudinal
displacements. This anisotropy is lost at the length scales above polymer
persistence length. For DNA, $l_p \approx 50$nm is much smaller than typical
dimensions of FCS sampling volume ($\sim 500$nm) and the expressions implying
isotropic dynamics can be used in all of the dynamic range probed by FCS. For
actin filaments with $l_p \approx 17\mu m$ the situation is reversed and the
anisotropy of segmental motion has to be taken into account explicitly.

In the next section we derive the FCS expressions for anisotropic motion of stiff
filaments. Although the segmental motion is dominated by the transverse
component, for the sake of generality, we will derive the expressions which take
into account explicitly both transverse and longitudinal displacements. Then we
use these expressions to analyze the results of two sets experiments with
different labeling strategies: 1) Partially (or locally) labeled filaments
(obtained via polymerization of nonfluorescent monomeric actin on fluorescent
seeds), and 2) homogenously labeled filaments (obtained by polymerization from a
mixture of labeled and unlabeled actin monomers). The analysis of the
experimental data with the expressions appropriate to each of the cases gives
consistent results on temporal dependence of monomers MSD. Finally, the
experimental data are compared to theoretical predictions for the dynamics of
semiflexible chains. The results are in qualitative and \emph{quantitative}
agreement with the theories taking into account hydrodynamic interactions between
the polymer segments \citep{Granek97, Harnau96}.

\section*{Theory}
FCS technique \citep{Webb72, Elson74, Magde74} (reviewed in e.g.
\citep{Schwille03, Thompson02, RiglerElson, KrichevskyBonnet}) is based on
monitoring fluctuations $\delta I_{em}(t) = I_{em} - \langle I_{em} \rangle$ in
fluorescence emission $I_{em}(t)$ as fluorescence species diffuse in a spatially
restricted excitation field, formed typically with the help of confocal optical
scheme \citep{Rigler93}. The autocorrelation function $G(t)=\langle \delta
I_{em}(0) \delta I_{em}(t)\rangle$ of emission fluctuations reflects the kinetics
of motion of fluorescent sources.

In this section we adapt the general formalism of FCS to the case of dynamics of
linear stiff polymers and derive expressions relating FCS correlation function to
the temporal dependence of MSD $\msdt$ of polymer segments.

The instantaneous detected emission, average emission and the correlation
function of fluorescence fluctuations are found through the spatial distribution
$c(\vec{r},t)$ of fluorescent labels and excitation-detection profile
$I(\vec{r})$ \citep{Elson74}:
\begin{eqnarray}
    &I_{em}(t)=Q \int d\vec{r}I(\vec{r})c(\vec{r},t), \nonumber \\
    &\langle {I}_{em} \rangle =Q \bar{c} \int d\vec{r}I(\vec{r}), \label{Eq.Iave1}\\
    &G(t) = Q^2 \int
    d\vec{r}d\vec{r}\,' I(\vec{r})I(\vec{r}\,')\langle \delta c(\vec{r}, 0) \delta
    c(\vec{r}\,',t)\rangle,   \label{Eq.CFr}
\end{eqnarray}
where $Q$ is specific brightness of a fluorescent molecule dependent on
fluorophore properties and the efficiency of detection optics, $\bar{c}=\langle
c(\vec{r},t)\rangle$ is the average concentration of fluorophores, and $\delta
c(\vec{r},t) = c(\vec{r},t) - \bar{c}$. Following \citep{BernePecora}
Eq.~\ref{Eq.CFr} can be rewritten as:
\begin{equation}
    G(t) = \frac{(2\pi)^3Q^2}{V} \int
    d\vec{q} \,|\tilde{I}(\vec{q})|^2 \langle \delta \tilde{c}^{*}(\vec{q}, 0) \delta
    \tilde{c}(\vec{q},t)\rangle,   \label{Eq.CFq}
\end{equation}
where "tilde" denotes spatial Fourier transform of the corresponding quantities,
such as e.g. $\tilde{I}(\vec{q})=(2\pi)^{-3/2}\int
d\vec{r}\,I(\vec{r})e^{-i\vec{q}\vec{r}}$, and $V$ is the total volume of the
sample.

We assume that fluorescent molecules are not moving independently, but are
attached to relatively large objects which themselves move independently of each
other. We will further assume that the objects are statistically equivalent and
that the distribution of  fluorophores within each object with respect to its
center of mass is described by $\Phi(\vec{r},\sit)$, where $\sit$ denotes the set
of internal degrees of freedom describing the current conformation and the
orientation of the object. Then $c(\vec{r}, t)=\sum_{j}\Phi\left
(\vec{r}-\vec{r}_j(t), \sitj \right)$, $\tilde{c}(\vec{q},
t)=\sum_{j}\tilde{\Phi}\left (\vec{q}, \sitj \right)e^{-i\vec{q}\vec{r}_j(t)}$,
where $\vec{r}_j(t)$ and $\sitj$ define respectively the center-of-mass position
and internal conformation of object $j$ at time $t$. Finally, substituting these
formulas in Eq.~\ref{Eq.CFq} we have:
\begin{equation}
    G(t) = (2\pi)^3Q^2 \bar{n} \int
    d\vec{q} \,|\tilde{I}(\vec{q})|^2 \langle \tilde{\Phi}^{*}\left (\vec{q},
    \siz \right ) \tilde{\Phi}\left (\vec{q}, \sit\right )
    e^{-i\vec{q}\Delta\vec{r}(t)}\rangle,
    \label{Eq.CFq2}
\end{equation}
where $\bar{n}$ is the average concentration of the objects (number of objects
per unit volume), $\Delta \vec{r}(t)= \vec{r}_j(t)- \vec{r}_j(0)$ is the
displacement of an object, and the index $j$ was omitted everywhere in
Eq.~\ref{Eq.CFq2} due to the statistical equivalence of the objects.

In its general form Eq.~\ref{Eq.CFq2} can be applied to any objects which have
some internal structure and internal dynamics.

We assume now that the objects are uniformly labeled segments of semi-flexible
polymers (one labeled segment per polymer, which, in general, are longer than
their labeled parts and contain unlabeled parts). For sufficiently stiff
polymers, we can neglect the dynamics within the labeled segments and consider
them to be straight. In this case the set of internal degrees of freedom
$\hat{s}$ reduces to a unit vector $\vec{s}$ defining the orientation of the
segment. Within the same approximation we can assume that any given labeled
segment moves without change in its orientation $\vec{s}_j(0)=\vec{s}_j(t)$. We
will discuss in detail the validity of our assumptions at the end of this
section. Here we just note that although these assumptions may look prohibitively
restrictive, for sufficiently stiff filaments they are valid in a wide range of
segmental displacements.

For a thin straight segment of length $L$ uniformly labeled with linear density
$\sigma$ of fluorophores:
\begin{equation}
    \tilde{\Phi}\left (\vec{q},\vec{s}\right) = \frac{\sigma}{\pi\sqrt{2\pi}} \frac{\sin
(\frac{1}{2}\,\vec{q}\,\vec{s}\,L)}{\vec{q}\,\vec{s}}=\frac{\sigma}{\pi\sqrt{2\pi}}
\frac{\sin (\frac{1}{2}\,qL\cos \alpha)}{q\cos \alpha},
\label{Eq.Phi}
\end{equation}
where $\alpha$ is an angle between $\vec{q}$ and $\vec{s}$. Furthermore, we split
the segmental displacement into components parallel and perpendicular to the
segment $\Delta \vec{r} = \Delta \vec{r}_{\|} + \Delta \vec{r}_{\bot}$, we assume
the two components to be independent of each other and to be Gaussian random
variables. Then, for fixed $\vec{q}$ and $\vec{s}$ the ensemble average of
$e^{-i\vec{q}\Delta\vec{r}(t)}$ is given by:
\begin{equation}
    \langle e^{-i\vec{q}\Delta\vec{r}(t)} \rangle_{|\vec{q},\vec{s}}=\exp
    \left ( -\frac{q^2 \cos^2 \alpha}{2} \langle \Delta r_{\|}^2\rangle
    -\frac{q^2 \sin^2 \alpha}{4} \langle \Delta r_{\bot}^2\rangle \right )
    \label{Eq.exp}
\end{equation}
The difference in the numeric prefactors in the two terms in the Eq.~\ref{Eq.exp}
stems from the fact that $\Delta \vec{r}_{\|}$ is defined on a line (parallel to
$\vec{s}$), while $\Delta \vec{r}_{\bot}$ is defined in a plane (perpendicular to
$\vec{s}$).

We assume, as it is usually done for confocal setups, the excitation-detection
profile $I(\vec{r})$ to be 3D Gaussian axisymmetric with respect to optical axis
$Z$:
\begin{eqnarray}
    &I(\vec{r})=I_0 \exp \left( -\frac{2(x^2+y^2)}{w_{xy}^2}
    -\frac{2z^2}{w_{z}^2}\right) \label{Eq.Ir} \\
    &\tilde{I}(\vec{q})=\frac {I_{0} w_{xy}^2 w_z}{8} \exp
    \left( -\frac{w_{xy}^2}{8}q^2\sin^2\theta
    -\frac{w_{z}^2}{8}q^2\cos^2\theta\right),
    \label{Eq.Iq}
\end{eqnarray}
where $w_{xy}$ and $w_z$ define the width of the profile in the $XY$ plane and in
$Z$ direction respectively, and $\theta$ is an angle between $\vec{q}$ and
Z-axis.

Substituting (\ref{Eq.Phi}),(\ref{Eq.exp}) and (\ref{Eq.Iq}) into
Eq.~\ref{Eq.CFq2}, averaging over $\alpha$ for a given $\vec{q}$ and integrating
over $\vec{q}$, we have:
\begin{equation}
    G(t) = \frac{\pi}{16}I_{0}^2 w_{xy}^5 \omega^2 Q^2 \sigma^2 \bar{n}
    \int_{0}^{\infty}dk \int_{-1}^{1}du \int_{-1}^{1}\frac{dp}{p^2}\sin^2 \left(
    \frac{\lambda kp}{2}\right) \exp \left( -\frac{k^2}{4}f(h_{\bot}^2, h_{\|}^2, p,
    u)\right),
    \label{Eq.Gt}
 \end{equation}
where $u=\cos \theta$, $p=\cos \alpha$, reduced units $k=qw_{xy}$, $\lambda =
L/w_{xy}$, $h_{\bot}^2 = \langle \Delta r_{\bot}^2\rangle /w_{xy}^2$, $h_{\|}^2 =
\langle \Delta r_{\|}^2\rangle /w_{xy}^2$ and $\omega = w_z/w_{xy}$ were
introduced, and $f$ denotes the following expression:
\begin{equation}
   f(h_{\bot}^2, h_{\|}^2, p,
    u)=h_{\bot}^2-p^2(h_{\bot}^2-2h_{\|}^2)+u^2(\omega^2-1)+1
\end{equation}

FCS correlation function is usually normalized by the square of the average
emission $G_1(t)=\langle \delta I_{em}(0) \delta I_{em}(t)\rangle/ \langle I_{em}
\rangle^2$. With this normalization the amplitude of the correlation function at
short time scales is the inverse of the average number $N$ of molecules in the
detection volume (given by $\pi^{3/2}w_{xy}^2 w_z$): $G_1(t\rightarrow 0)=1/N$.
Here we prefer another normalization: $G_2(t)=\langle \delta I_{em}(0) \delta
I_{em}(t)\rangle/ \langle I_{em} \rangle = \langle I_{em} \rangle G_1(t)$. A
correlation function defined this way is independent of the concentration of the
moving species (as long as there are no interactions between the objects) and its
amplitude at short time scales gives the fluorescence per moving object:
$G_2(t\rightarrow 0)=\langle I_{em} \rangle/N$. This is an interesting quantity
in the context of labeled segments comparable or larger than $w_{xy}$: in this
case only part of the labeled segment can "fit" into the detection volume and
contribute to the correlation function. The length of this part can be estimated
from $G_2(t\rightarrow 0)$.

In order to calculate $G_2(t)$ we substitute $\bar{c}=\bar{n}\,\sigma L$ and
(\ref{Eq.Ir}) into (\ref{Eq.Iave1}) to find
\begin{equation}
    \langle I_{em}\rangle = (\pi/2)^{3/2}I_0 w_{xy}^2 w_z Q \bar{n}\sigma L.
    \label{Eq.Iave2}
\end{equation}
Performing integration over $k$ in (\ref{Eq.Gt}) and making use of
(\ref{Eq.Iave2}) and of definition of $G_2$ we obtain:
\begin{equation}
    G_2(t)=\frac{I_0 Q}{2\sqrt{2}}\sigma w_{xy} \frac{\omega}{\lambda} \int_0^1
    du \int_0^1 dp \, \frac{1-\exp(-\lambda^2 p^2/f(h_{\bot}^2, h_{\|}^2, p,
    u))}{p^2\sqrt{f(h_{\bot}^2, h_{\|}^2, p, u)}}.
    \label{Eq.G2gen}
\end{equation}
Given the knowledge of the experimental geometry (i.e. parameters $w_{xy}$, $w_z$
and $L$), Eq.~\ref{Eq.G2gen} can be used to numerically calculate the relation
between the FCS correlation function and temporal behavior of segmental MSDs,
i.e. $h_{\bot}^2(t)$ and $h_{\|}^2(t)$.

In two important limiting cases the explicit expressions can be derived for
$G_2(t)$: 1) for very short labeled segments, i.e. $L \rightarrow 0$ while
$L\sigma = const$, and 2) for very long labeled segments, i.e. $L \rightarrow
\infty$ while $\sigma = const$.

In the case of short segments, we obtain:
\begin{equation}
G_2(t)=\frac{I_0 Q}{2\sqrt{2}} \frac{\sigma
L\omega}{\sqrt{(1+h_{\bot}^2)(\omega^2-1)(h_{\bot}^2-2h_{\|}^2)}} \ln
\frac{\sqrt{(1+h_{\bot}^2)(\omega^2+2h_{\|}^2))}+\sqrt{(\omega^2-1)(h_{\bot}^2-2h_{\|}^2)}}
{\sqrt{(1+2h_{\|}^2)(\omega^2+h_{\bot}^2)}} \label{Eq.Point}
\end{equation}

We note that for isotropic motion, i.e. for $h_{\bot}^2=2h_{\|}^2=\langle \Delta
r^2 \rangle /(3w_{xy}^2) $, the Eq.~\ref{Eq.Point} reduces to the more standard
FCS expression for the random motion of point-like objects:
\begin{equation}
    G_2(t)= \frac{QI_0}{2\sqrt{2}}(\sigma L) \left(1+\frac{2}{3}\frac{\langle
    r^{2}(t)\rangle}{w_{xy}^{2}}\right)^{-1}\left(1+\frac{2}{3}\frac{\langle
    r^{2}(t)\rangle}{w_{z}^{2}}\right)^{-1/2}.
    \label{eq1}
\end{equation}

For infinitely long segments Eq.~\ref{Eq.G2gen} gives:
\begin{equation}
    G_2(t)=\frac{QI_0}{2\sqrt{2}}\frac{\sigma w_{xy} \omega \sqrt{\pi}}
    {\sqrt{(\omega^2-1)(1+h_{\bot}^2)}} \arctan
    \sqrt{\frac{\omega^2-1}{1+h_{\bot}^2}}.
    \label{Eq.Inf}
\end{equation}
Note that $G_2(t)$ in this case is independent of $h_{\|}^2$ since longitudinal
motion of infinitely long labeled segments does not lead to any fluctuations in
fluorescence.

Although unrelated to our experiments, still an interesting particular case of
application of Eq.~\ref{Eq.Inf} is that of spherically symmetric detection volume
$w_{xy} =w_z=w$. In this case $G_2(t)=\sqrt{\pi/8}\,QI_0w\sigma
(1+h_{\bot}^2)^{-1}$ is similar to the correlation function produced by a planar
motion of point-like objects.

We return now to Eq.~\ref{Eq.G2gen} in order to find the dependence of the
fluorescence per moving object $G_2(t \rightarrow 0)$ on the length of the
labeled segment. We make use of the fact that for small segments $L \ll w_{xy}$:
$G_2(t \rightarrow 0) = QI_0 \sigma L/(2\sqrt{2})$, and define an apparent length
$L_{app}$ and respectively $\lambda_{app}=L_{app}/w_{xy}$ such that for the
labeled segment of any length $G_2(t \rightarrow 0) = QI_0 \sigma
L_{app}/(2\sqrt{2})$.

For $t = 0$, i.e. $h_{\|}^2=h_{\bot}^2=0$ it is possible to perform integration
over $u$ in Eq.~\ref{Eq.G2gen} and arrive to the expression relating the apparent
length of the labeled segment to its actual length and to the parameters of the
detection volume:
\begin{equation}
    \lambda_{app}=\frac{\omega}{\lambda \sqrt{\omega^2-1}}\left[ -\ln(\omega+\sqrt{\omega^2-1})
    +\int_1^\omega \frac{dv}{v\sqrt{v^2-1}}\left( \lambda \sqrt{\pi}\, \mathrm{erf}
    \left(\frac{\lambda}{v}\right) + v e^{-\lambda^2/v^2}\right) \right],
    \label{Eq.Lapp}
\end{equation}
where $\mathrm{erf}(x)=2\pi^{-1/2}\int_0^x e^{-t^2}dt$

An example of $\lambda_{app}(\lambda)$ dependence is shown in Fig.~\ref{Fig:Lapp}
for $\omega = 5$ corresponding to our experimental geometry. As expected for
$\lambda < 1$, all of the segment can fit into the detection volume and
$\lambda_{app} \approx \lambda)$. For $\lambda \gg 1$ the apparent length
saturates at the value which can be found from (\ref{Eq.Inf}):
\begin{equation}
    \lambda_{app}(\lambda \rightarrow \infty)= \omega
    \sqrt{\frac{\pi}{\omega^2-1}} \arctan \sqrt{\omega^2-1}.
\end{equation}

Finally, we discuss the validity of assumptions leading to Eq.~\ref{Eq.G2gen}.
The main assumption we made was to neglect the internal dynamics within the
labeled segment. This assumption in fact just puts a lower limit on the
accessible range of studied segmental displacements: the derived equations are
valid as long as the center-of-mass motion $\Delta\vec{r}(t)$ of the segment is
larger than the characteristic motions \emph{within} the segment.

The characteristic internal motions $\langle \Delta r_{int} ^2\rangle$ within the
labeled segment can be estimated to be of the order of $\langle \Delta r_{int}
^2\rangle = (2/45)L^{3}/l_p$ \citep{Granek97}. For segmental displacements larger
than that, motions \emph{within} the labeled part can be neglected and the
labeled segment can be considered essentially rigid and moving as a single unit.

This condition seemingly prohibits studies with long labeled segments, which have
considerable motions within them. However, only a small part of the filament (of
length $\approx L_{app} < L$) can cross the sampling volume and contribute to the
fluctuations in fluorescence at any given moment. Thus the lower limit for the
range of accessible segmental motions can be further relaxed to
$(2/45)L_{app}^{3}/l_p$. E.g. even for very long homogeneously labeled actin
filaments ($L \gg w_{xy}$, $L_{app} \approx 2.5 w_{xy}$) the segmental dynamics
can be studied in the range $\langle \Delta r ^2\rangle > 4\cdot 10^{-4} \mu
m^2$.

The expressions derived in this section (such as Eqs.~\ref{Eq.G2gen},
\ref{Eq.Point}, \ref{Eq.Inf}) can be directly used to measure segmental
displacements from FCS correlation functions: only the parameters defining the
experimental geometry ($w_{xy}$, $w_z$ and in the case of Eq.~\ref{Eq.G2gen} $L$
in addition) need to be calibrated. All other parameters affect $G_2(t
\rightarrow 0)$, the value of which can be determined from the plateau level of
the experimentally measured correlation function at short time-scales. Finally,
for the case of internal dynamics of a semi-flexible chain, relevant to most of
the studies on actin dynamics as well as to our studies presented here, the
longitudinal displacement $h_{\|}^2$ can be neglected in comparison to transverse
motion $h_{\bot}^2$. Then, with known geometrical parameters and measured $G_2(t
\rightarrow 0)$, Eq.~\ref{Eq.G2gen} and its limiting cases Eqs.~\ref{Eq.Point}
and \ref{Eq.Inf} give a one-to-one relation between segmental MSD and FCS
correlation function. Some examples of such dependence are given in
Fig.~\ref{Fig:Corr-Theory}. Using these dependencies, the experimentally measured
correlation function $G_2(t)$ can be converted into the temporal dependence of
segmental MSD $\langle \Delta r_{\bot}^2(t)\rangle$ ($=w_{xy}^2 h_{\bot}^2(t)$).


\section*{Materials and Methods}
\subsection*{Actin preparation}
Unlabeled actin is purified from chicken skeletal muscle acetone powder and
stored in G-buffer \citep{Pardee82}. Two sets of samples are prepared: 1)
filaments labeled at defined positions, and 2) homogenously labeled filaments.

In order to label the filaments at defined positions (sample 1), we utilize
fluorescently labeled actin filament seeds as templates for additional
polymerization of unlabeled actin monomers. To prepare fluorescent seeds we first
polymerize 1$\mu$M of fluorescent G-actin (Actin Alexa 568, Molecular Probes,
Eugene, OR,  or Rhodamine-Actin, Cytoskeleton, Denver, CO) in the presence of
phalloidin (Molecular Probes) to stabilize the filaments (1:1 actin to phalloidin
molar ratio). The labeled filaments are then broken by a brief sonication and
vigorous pipetting into short fragments (average length of $\sim 170$nm,
estimated by FCS see \textbf{Results and Discussion}). These fragments are then
used as seeds for further polymerization of unlabeled G-actin (9$\mu$M).
Polymerization proceeds for 10 min at room temperature. We note that concurrent
with polymerization, there is an ongoing annealing process \citep{Pollard01}
which both increases polymer length and creates multiply labeled filaments.
Independent fluorescence microscopy observation confirms that at the beginning of
the experiment about ten percents of filaments are labeled at two distinct
positions (typically separated by more than $1\mu$m), while the rest are
single-labeled. Since the distance between the labeled portions of the
double-labeled filaments is much larger than confocal radius (0.21 $\mu$m), the
"cross-talk" between the labels can be neglected and the formalism derived in the
preceding section can be applied.

The homogenously labeled filaments (sample 2) are prepared in a similar manner.
The difference is that the seeds are prepared from a mixture of labeled and
unlabeled actin monomers (1:9 molar ratio) and further polymerization proceeds
with the same mixture.

This procedure results in actin filaments of several microns in length: $\sim
4\mu m$ on average at the beginning of the experiment and growing in the course
of experiment to $\sim 8\mu m$ due to annealing, as verified by fluorescence
microscopy.

For the experiments, the solution is diluted tenfold to a final actin
concentration of 1$\mu$M. Typically, one microliter of solution is sealed between
two glass coverslips separated by a 250$\mu$m spacer. To prevent protein
adsorption, the glass coverslips are coated with an inert polymer (Polyethylene
Glycol) according to the protocol of \citep{Nassoy02}. Most of the measurements
were carried out at a distance of 40$\mu$m from the surface. This distance was
chosen on purpose to be larger than F-actin length in order to minimize the
effect of surface proximity on filament dynamics. Control experiments performed
at a distance of $100\mu$m from the surface give results identical to those
presented here.

The experiments are started immediately after dilution and are conducted within
30 minutes. The main reason to limit the duration of experiment is to minimize
the effect of actin filament fragmentation, which leads to appearance of short
filaments and associated noise in the FCS correlation function. After $\sim$1
hour of FCS measurement we start to see the changes in the MSD due to the
fragmentation: the MSD vs. time curve shifts to larger displacements. To be on
the safe side, we limit all of our measurements to the first 30 min after
polymerization.

\subsection*{Experimental setup}
The optical setup is home-built based on the Nikon Eclipse TE300 inverted
microscope (Nikon Corporation, Tokyo, Japan). The confocal excitation is provided
by 514nm line ($\sim 2.5\mu$W power before microscope objective) of an Ar-ion
laser (Advantage 163D, Spectra-Physics, Mountain View, CA) deflected by Q525
dicroic beamsplitter (Chroma Technology, Rockingham, VT) into a high-power
objective lens (UPLAPO 60X1.2W, Olympus Europe, Hamburg, Germany). The collected
emission passes through the beamsplitter, then a bandpass filter HQ565/80 (Chroma
Technology, Rockingham, VT) and a pinhole of $50\mu m$ in diameter. The emission
is detected by a photon counting avalanche photodiode (SPCM-AQR-14 PerkinElmer
Optoelectronics, Vaudreuil, Quebec, Canada) whose output is fed into digital
correlator Flex2k-12Dx2 (Correlator.com, Bridgewater, NJ). The correlator is
capable of working in two modes, either as traditional correlator carrying out
the correlation analysis of emission online, or as photon history recorder,
storing the time arrivals of every photon on computer hard drive. For the
presented experiments, we make use of the photon history recorder mode, while
analyzing the recorded photon traces offline with software correlator as
described in \textbf{Data Analysis}. The correlator program was written as
C-module running under MATLAB environment (MathWorks, Natick, MA).

The parameters $w_{xy}\approx 0.21\mu m$, $w_{z}\approx 1.1 \mu m$ of the
confocal volume are calibrated before and after each experiment by measuring the
diffusion of free Rh6G fluorophores \citep{Rigler93}.

\subsection*{Data Analysis}
\begin{figure}
   \begin{center}
      \includegraphics*[width=3.25in]{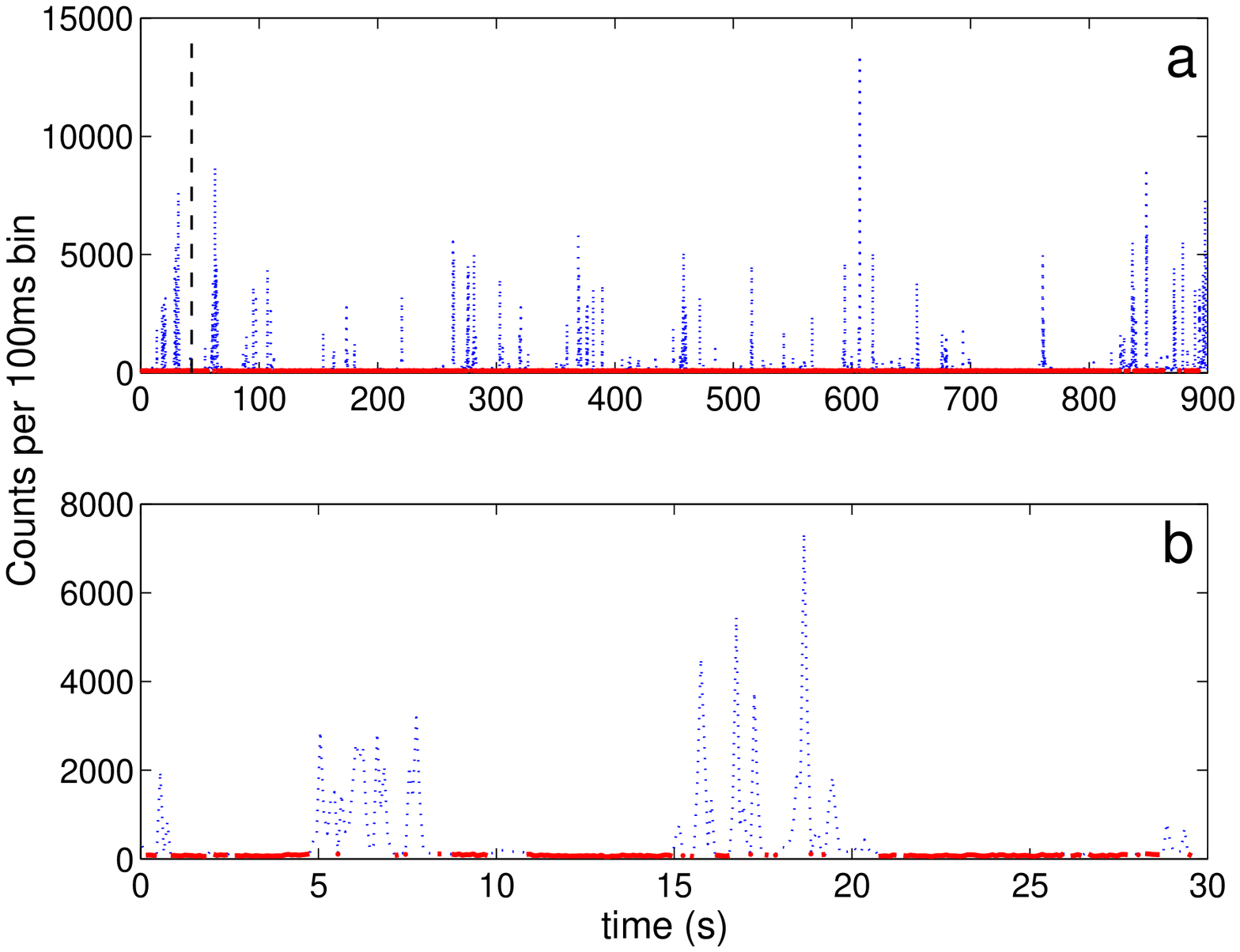}
      \caption{}
      \label{Fig:trace}
   \end{center}
\end{figure}

\emph{Sample 1:} The measurement of the photon emission count rate from locally
labeled F-actin reveals that photons arrive in intense bursts of $\sim 10^{5}$
counts/sec lasting $\sim$ 0.1 to 1s, separated by intervals of low count rate of
$\sim 10^3$ counts/sec (Fig.~\ref{Fig:trace}). The bursts are caused by the
passage of the labeled F-actin through the confocal volume. The fluorescence in
between the bursts originates from residual free fluorophores diffusing in the
sample. The motion of the free fluorophores results in a correlated background
noise $I_{b}(t)$ which adds up with the labeled actin signal $I(t)$ to a total
emission $I_{tot}=I+I_{b}$. Thus the overall correlation function
$G_{tot}(t)=\langle \delta I_{tot}(0) \delta I_{tot}(t)\rangle$ is given by:
\begin{equation}
    G_{tot}(t)= G(t)+G_{b}(t),
    \label{eq2}
\end{equation}
where $G(t)=\langle \delta I(0) \delta I(t)\rangle$ and $G_{b}(t)=\langle \delta
I_{b}(0) \delta I_{b}(t)\rangle$ are the correlation functions of emission from
labeled segments and from free fluorophores respectively.

Although the overall contribution of free fluorophores $G_{b}$ to the total
correlation function is small, their fast motion is responsible for most of the
decay of the correlation function at the time scales below 1ms, where the motion
of the filament segments is negligible. This could limit the analysis of
monomers' MSD to the time scales above 1ms. However, as shown below, it is
possible to separate the contributions of labeled actin and free fluorophores
within the \emph{same} experiment, and thus, extend the range of measurements to
time scales as low as $\sim 40\mu$s.

In order to separate free fluorophore noise from the signal, we record the
complete photon trace, i.e. times between arrivals of consecutive photons (with
temporal resolution of 16.7ns). The photons are then binned into 100ms intervals
$\{I(t_n)\}$ (Fig.~\ref{Fig:trace}). The stretches of time with no bursts are
determined and the background correlation function $G_{b}(t)$ is calculated on
these stretches using the original photon traces. The total correlation function
$G_{tot}(t)$ is computed using the complete photon trace. Finally, the
contribution of labeled actin $G(t)$ is obtained using Eq.~(\ref{eq2}). The
intensity-normalized correlation function of actin segments is found by $G_2(t) =
G(t)/(I_{tot}-I_{b})$.

We find this procedure more robust than the correlation analysis of bursts
intervals. First, there is some background contribution within the bursts as
well. Second, unlike bursts, the intervals of pure background can be determined
unambiguously: any intervals suspect to contain a burst can be deselected.

\begin{figure}
   \begin{center}
      \includegraphics*[width=3.25in]{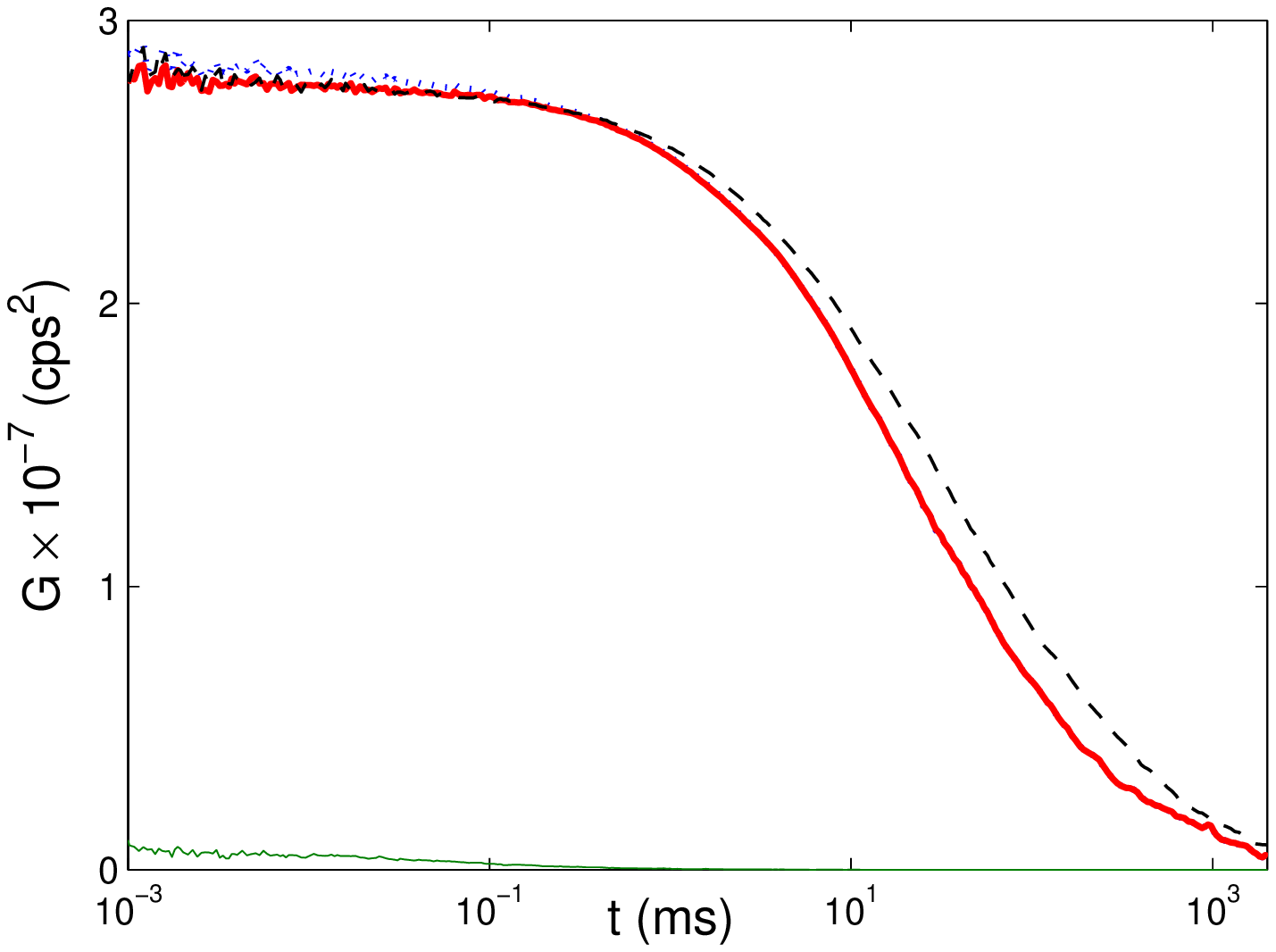}
      \caption{}
      \label{Fig:Corr-Exp}
   \end{center}
\end{figure}

Practically, the background is deduced by calculating the median intensity over
all bins $I_{med}=median(\{I(t_n)\})$, and selecting all of the intervals which
deviate from the median by less than $\sigma=median(\{(I(t_n)-I_{med})^2\})$ for
analysis of background noise. An example of this procedure is presented in
Fig.~\ref{Fig:trace}. The resulting total, background and signal correlation
functions $G_{tot}(t)$, $G_b(t)$ and $G(t)$ are shown in Fig.~\ref{Fig:Corr-Exp}.
The background correlation function $G_b(t)$ is indeed well described by 3D
diffusion model of the free fluorophore \citep{Rigler93} with a characteristic
decay time of 70$\mu$s. We note that the amplitude of $G(t)$ is much larger than
that of $G_b(t)$ mainly due to the multiple labeling of the filament segments:
thus the motion of the labeled segments leads to much larger fluctuations in
emission than the motion of single fluorophore molecules.

For calculation of monomers' MSD the amplitude of the correlation function
$G_0=G(t \rightarrow 0)$ is estimated from the level of the correlation function
in 3 to 20 $\mu$s range, which is above the characteristic time scales of the
fluorophore triplet state kinetics and below the characteristic time scales of
segmental dynamics.

\emph{Sample 2:} We use the same approach to analyze data from homogeneously
labeled F-actin. However, since the passage of the labeled segments through the
sampling volume is more frequent in this case as compared to Sample 1, we make
use of shorter binning intervals of 30ms and we pick the intervals with the
average count rate not exceeding $I_{med}+0.25\sigma$ for the analysis of
background noise. These conditions give a noise correlation function with
characteristic decay time below 100$\mu s$. The above parameters were found to be
optimal between less restrictive conditions which lead to a notable contribution
of the signal in the estimated noise correlation function (characterized by decay
times exceeding 1ms), and more restrictive parameters which clearly underestimate
the noise level.

\section*{Results and Discussion}

We present the correlation functions obtained from both types of samples in
Fig.~\ref{Fig:Corr-Exp}. To facilitate the comparison of temporal kinetics, the
amplitude of the correlation function of homogeneously labeled actin was adjusted
to the level of the correlation function of the partially labeled sample. The
functions look similar but are notably shifted in time: the correlation function
of the homogeneously labeled sample decays slower by a factor of $\sim 1.4$ than
that of the locally labeled F-actin.

\begin{figure}
   \begin{center}
      \includegraphics*[width=3.25in]{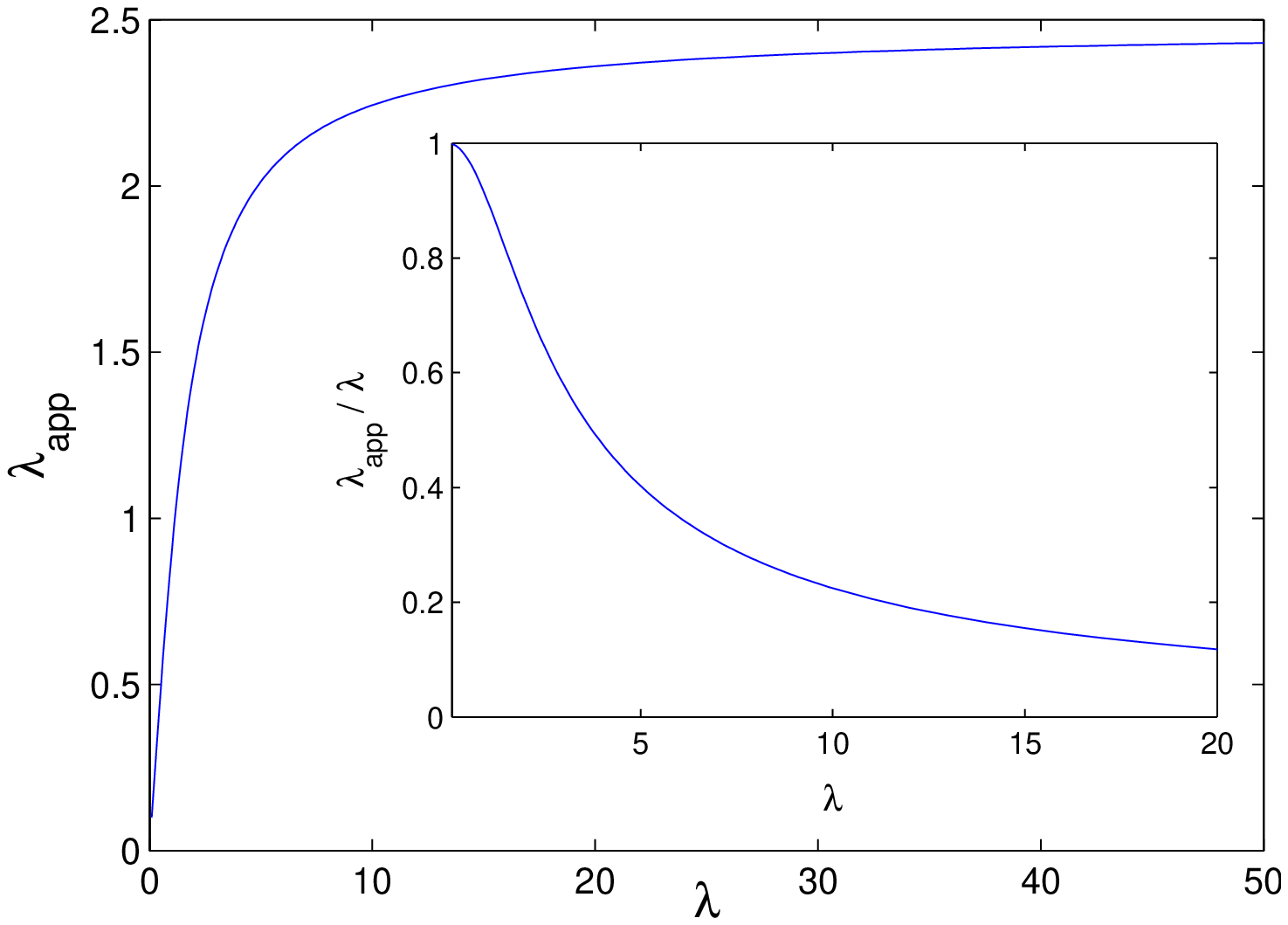}
      \caption{}
      \label{Fig:Lapp}
   \end{center}
\end{figure}

We can estimate the length of the labeled parts of Sample 1 by analyzing the
amplitude of its intensity-normalized correlation function $G_2(t\rightarrow 0)$.
$G_2$ amplitude is $62\pm 8$ times higher than the corresponding amplitude of the
correlation function of G-actin monomers obtained in similar conditions (data not
shown). Since 14 actin monomers form a filament of 37nm, this gives the apparent
length of the labeled segment of $L_{app}\approx 160 \pm 20 nm$. Converting
$L_{app}$ into real segment length (Eq.~\ref{Eq.Lapp} or Fig.~\ref{Fig:Lapp}) we
obtain $L \approx 170 \pm 30 nm$.

The apparent length of \emph{homogeneously} labeled F-actin is $\sim 470nm
\approx 2.3 w_{xy}$, as estimated from the amplitudes of the corresponding
correlation functions. This value is in a good agreement with our expectations
for long homogenously labeled filaments (Eq.~\ref{Eq.Lapp} and
Fig.~\ref{Fig:Lapp}).

\begin{figure}
   \begin{center}
      \includegraphics*[width=3.25in]{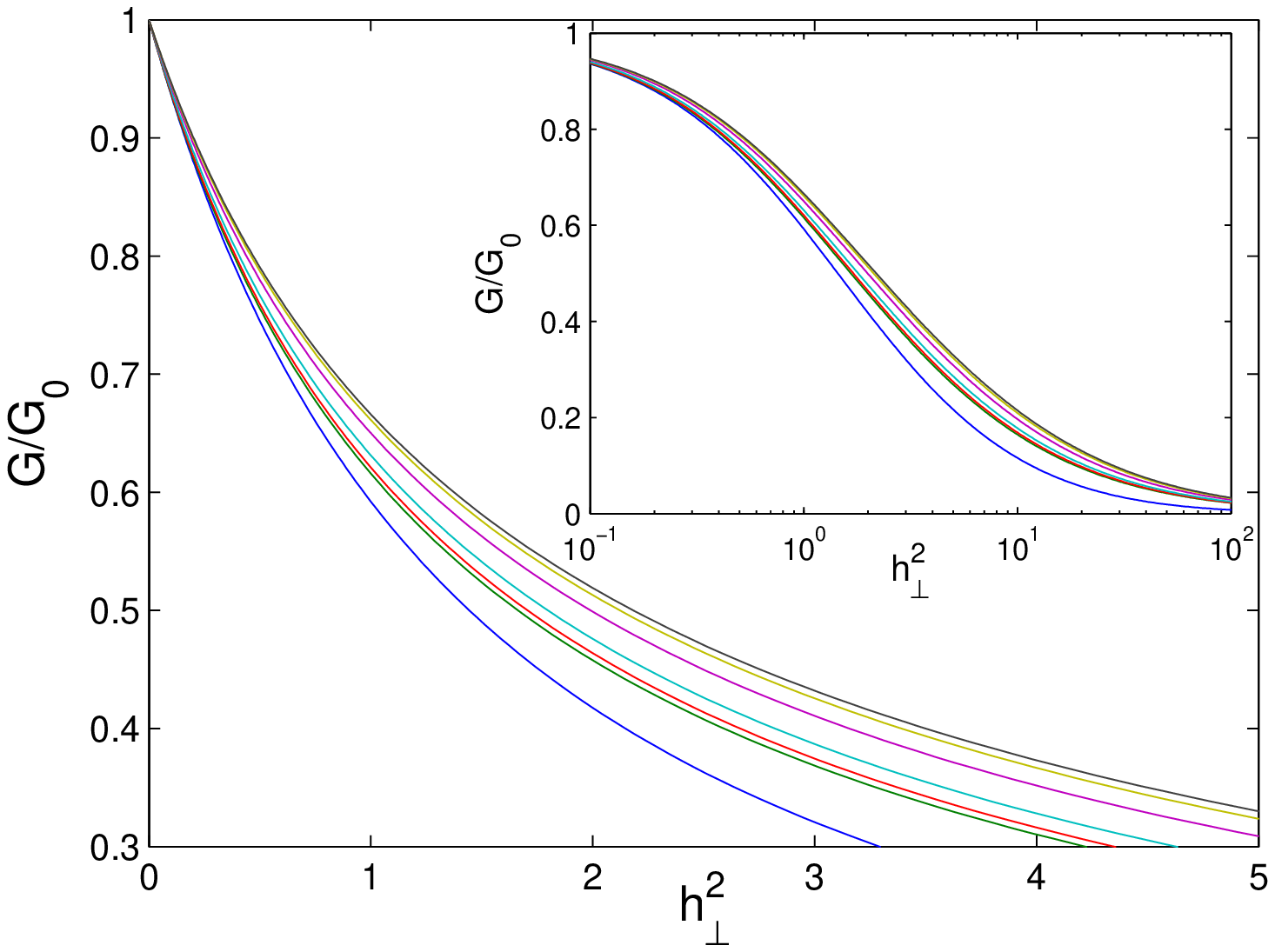}
      \caption{}
      \label{Fig:Corr-Theory}
   \end{center}
\end{figure}

As discussed above, in order to extract $\langle \Delta r_{\bot}^2(t)\rangle$
dependence from the correlation functions we neglect longitudinal motion of the
segments. Although the general expression Eq.~\ref{Eq.G2gen} can be used to
analyze both sets of data, the numerical calculation shows that for $L<w_{xy}$
(i.e. $\lambda < 1$) Eq.~\ref{Eq.G2gen} leads to essentially the same dependence
of the correlation function $G_2$ on MSD as Eq.~\ref{Eq.Point} derived for point
sources (Fig.~\ref{Fig:Corr-Theory}). Similarly, for homogeneously labeled
filaments with length larger than 3$\mu m$ ($\lambda > 15$), the Eq.~\ref{Eq.Inf}
for infinitely long labeled segment can be used (compare curves for $\lambda >
15$ and $\lambda
 \rightarrow \infty$ in Fig.~\ref{Fig:Corr-Theory}). Thus we make the use of explicit expressions
 Eq.~\ref{Eq.Point} and Eq.~\ref{Eq.Inf} to analyze data on partially and
homogeneously labeled polymers respectively. Practically, in both cases we
tabulate $G_2(h^2_{\bot})/G_2(0)$ for a wide range of $h^2_{\bot}$ as exemplified
in Fig.~\ref{Fig:Corr-Theory} (curves for $\lambda =0$ and $\lambda
 \rightarrow \infty$). We normalize the experimentally obtained correlation
 functions $G_2(t)$ by their plateau values. We use tabulated $G_2(h^2_{\bot})/G_2(0)$ dependencies to find
 $h^2_{\bot}$ values corresponding to each measured data point $G_2(t)$. This
 gives the $h^2_{\bot}(t)$ dependence (and respectively $\langle \Delta r_{\bot}^2(t)\rangle
 =w_{xy}^2 h^2_{\bot}(t)$).

The extracted temporal dependencies of transverse monomer motion are presented in
Fig.~\ref{Fig:Rsq}. Despite the difference in the temporal behavior of the
original correlation functions for partially and homogeneously labeled filaments,
the application of the appropriate expressions for each case leads to consistent
data on monomers dynamics for both sets of measurements. The data span a wide
range of time scales: from $\sim 40\mu s$ to $\sim 2s$.

In Fig.~\ref{Fig:Rsq} we compare the experimental data to the predictions of two
types of semi-flexible polymer dynamics theories, which do \citep{Granek97} and
do not account for the hydrodynamic interactions \citep{Harnau97}. The
measurements agree very well with the hydrodynamic theory over about four decades
in time. The deviation of the data at long time scales is probably related to the
diffusion of the filament as a whole.

To conclude, we adapt FCS formalism for the studies of the internal dynamics of
semi-flexible polymers. We make use of a developed formalism to obtain
non-invasive measurements on the kinetics of segmental motion in actin filaments.
Two labeling strategies, local labeling and homogeneous labeling, lead to
consistent results. The transverse segmental MSD  $\langle \Delta
r_{\bot}^2(t)\rangle$ is probed over a wide range of timescales, from $\sim
40\mu$s to $\sim 2$s. Almost over the whole range the data points follow closely
the prediction of hydrodynamic theories \citep{Granek97, Harnau96}. Thus,
non-invasive measurements of $\msdt$ carried out with FCS allow us to test
hydrodynamic theories directly over a wide temporal range. We note, finally, that
although in the presented measurements the transverse motion is the dominant mode
of motion, in other cases, such as F-actin in the presence of molecular motors
(myosins) the longitudinal motion may be significant. Then the two suggested
labeling strategies, partial and homogeneous labeling, being very different in
their sensitivity to longitudinal motion, can be used in conjunction to separate
transverse and longitudinal components of segmental displacement.

We are indebted to R. Granek, E. Frey and K. Kroy for fruitful discussions and
for valuable suggestions. We thank also D. Groswasser for careful reading of the
manuscript. This work has been supported by the Israel Science Foundation grants
No.229/01 and No.663/04.


\clearpage
\section*{Figure Legends}

\subsubsection*{Figure~\ref{Fig:Lapp}.}
Dependence of the apparent length of the labeled segment on its real length as
given by Eq.~\ref{Eq.Lapp} for $\omega=5$ (close to the aspect ratio of the
sampling volume in our setup). The lengths are given in the units of confocal
volume radius $\lambda = L/w_{xy}$, $\lambda_{app}=L_{app}/w_{xy}$ . The apparent
length of very long $\lambda \gg 1$ segments approaches $\approx 2.5w_{xy}$.
\emph{Inset:} Ratio of the apparent length to the real length of the labeled
segment. Labeled segment lengths of $L < w_{xy}$ ($\lambda < 1$) results in
$L_{app} \approx L$.

\subsubsection*{Figure~\ref{Fig:Corr-Theory}.}
Calculated dependencies of FCS correlation function on the transverse mean-square
displacement of semi-flexible polymer segment. FCS correlation functions are
normalized by their zero-time values (to have unit amplitude) and transverse MSD
is given in the units of confocal volume radius $h_{\bot}^2 = \langle \Delta
r_{\bot}^2\rangle /w_{xy}^2$. The leftmost curve is given by standard FCS
expression Eq.~\ref{eq1} for isotropic motion. The other curves are calculated
from Eqs.~\ref{Eq.G2gen}, \ref{Eq.Point}, \ref{Eq.Inf} for different values of
$\lambda$, \emph{left to right:} 0, 1, 2, 5, 15, $\infty$. \emph{Inset:} Same
curves in semilog scale allowing to assess wider range of $h_{\bot}^2$.

\subsubsection*{Figure~\ref{Fig:trace}.}
Photon count trace of partially labeled actin filaments (\emph{Sample 1}). The
photon trace is collected with 16.7ns resolution and split into 100ms bins. The
data points represent photon counts per bin: (a) Full trace over the duration of
one measurement, (b) Zoom into first 30s of measurement. Bursts (dotted line) due
to the passage of labeled parts are separated by the intervals of background
noise (solid line).

\subsubsection*{Figure~\ref{Fig:Corr-Exp}.}
Total $G_{tot}(t)$ (dotted line), background $G_b(t)$ (thin solid line) and
signal $G(t)$ (thick solid line) correlation functions of partially labeled
filaments (\emph{Sample 1}). $G_{tot}(t)$ is obtained by analyzing the complete
photon trace, $G_b(t)$ is caused by the diffusion of free fluorophore and is
calculated from the photon trace in between the bursts in Fig.~\ref{Fig:trace}.
$G(t)=G_{tot}(t)-G_b(t)$ is the correlation function resulting from the motion of
F-actin labeled segments after background substraction. Dashed line is the
correlation function of homogeneously labeled actin (\emph{Sample 2}) after
subtraction of background noise. The correlation function of \emph{Sample 2} was
normalized to have the same amplitude as that of \emph{Sample 1} in order to
facilitate the comparison of temporal behavior.

\begin{figure}
   \begin{center}
      \includegraphics*[width=3.25in]{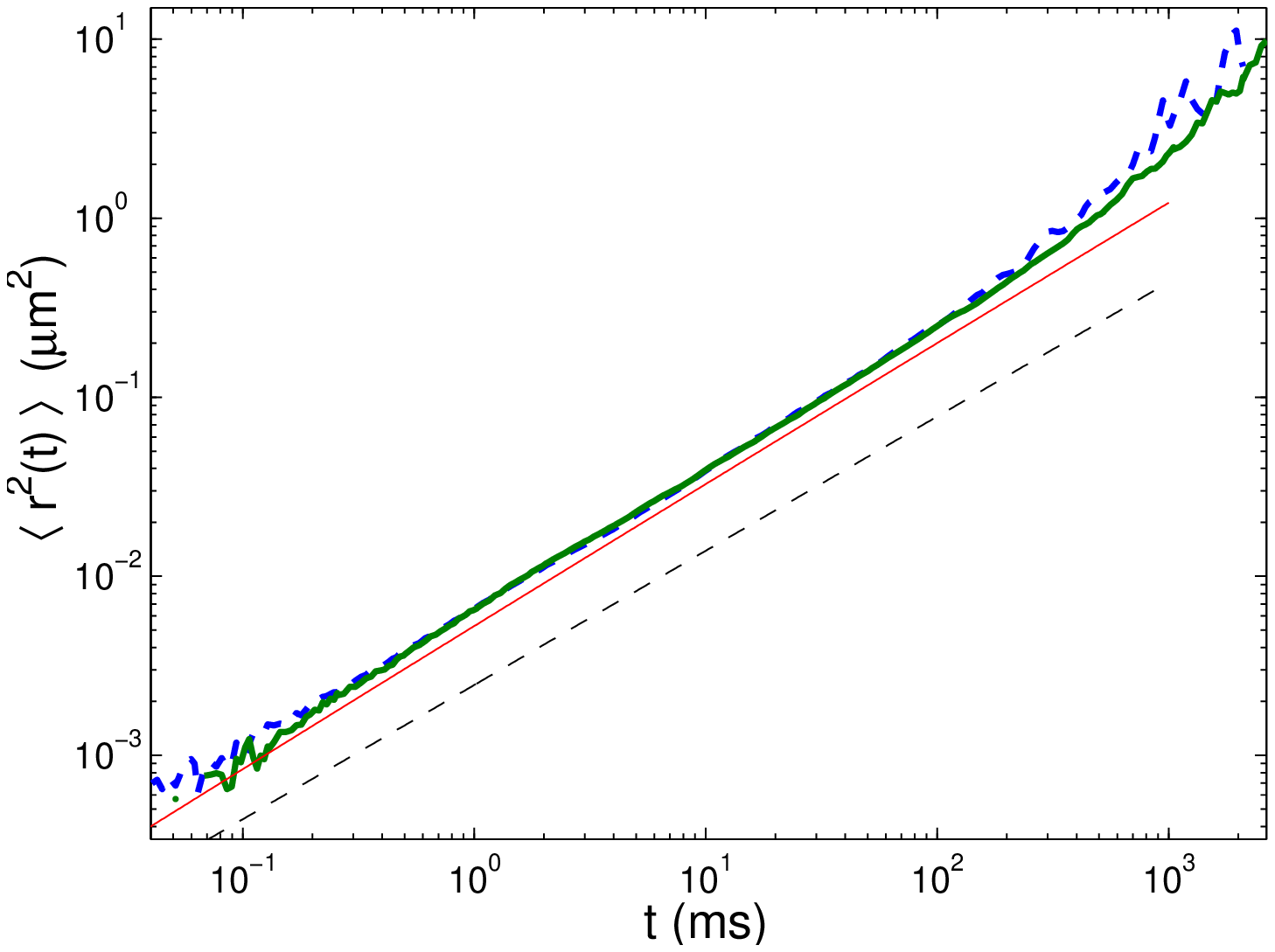}
      \caption{}
      \label{Fig:Rsq}
   \end{center}
\end{figure}

\subsubsection*{Figure~\ref{Fig:Rsq}.}
The kinetics of random motion $\langle \Delta r_{\bot}^2(t)\rangle$ of actin
filaments' segments. Experimental measurements on locally labeled (thick dashed
line) and on homogeneously labeled (thick solid line) are compared to the
theoretical predictions by the hydrodynamic theory of Granek \citep{Granek97}
(thin solid line) and non-hydrodynamic theory by Harnau et al \citep{Harnau96}
(thin dashed line). The parameters used for the calculation are $l_p=17 \mu m$,
filament diameter of $7nm$, solvent viscosity 1$mPa \cdot s$, and filament length
of 6$\mu m$.

\clearpage

\clearpage

\clearpage

\clearpage

\clearpage

\end{document}